# Evolution of ferroelectric properties in $Sm_xBi_{1-x}FeO_3$ via automated Piezoresponse Force Microscopy across combinatorial spread libraries

*Aditya Raghavan,[1] Rohit Pant,[2] Ichiro Takeuchi,[2] Eugene A. Eliseev,[3] Marti Checa,[4] Anna N. Morozovska,[5,*] Maxim Ziatdinov,[6] Sergei V. Kalinin,[1,6,†] and Yongtao Liu[4,‡]*

[1] Department of Materials Science and Engineering, University of Tennessee, Knoxville, Tennessee 37909, United States of America

[2] Deparment of Materials Science and Engineering, University of Maryland, College Park, Maryland 20742, United States of America

[3] Frantsevich Institute for Problems in Materials Science, Omeliana Pritsaka str., 3, Ukraine, 03142, Kyiv, Ukraine

[4] Center for Nanophase Materials Sciences, Oak Ridge National Laboratory, Oak Ridge, Tennessee 37830, United States of America

[5] Institute of Physics, National Academy of Sciences of Ukraine, 46, pr. Nauky, 03028 Kyiv, Ukraine

[6] Physical Sciences Division, Pacific Northwest National Laboratory, Richland, Washington 99354, United States of America

Notice: This manuscript has been authored by UT-Battelle, LLC, under Contract No. DE-AC0500OR22725 with the U.S. Department of Energy. The United States Government retains and the publisher, by accepting the article for publication, acknowledges that the United States Government retains a non-exclusive, paid-up, irrevocable, world-wide license to publish or reproduce the published form of this manuscript, or allow others to do so, for the United States Government purposes. The Department of Energy will provide public access to these results of federally sponsored research in accordance with the DOE Public Access Plan (https://www.energy.gov/doe-public-access-plan).

[*] anna.n.morozovska@gmail.com

[†] sergei2@utk.edu

[‡] liuy3@ornl.gov

**Abstract**

Combinatorial spread libraries offer a unique approach to explore evolution of materials properties over the broad concentration, temperature, and growth parameter spaces. However, the traditional limitation of this approach is the requirement for the read-out of functional properties across the library. Here we demonstrate the application of automated Piezoresponse Force Microscopy (PFM) for the exploration of the physics in the $Sm_xBi_{1-x}FeO_3$ system with the ferroelectric-antiferroelectric morphotropic phase boundary. This approach relies on the synergy of the quantitative nature of PFM and the implementation of automated experiments that allows PFM-based gird sampling over macroscopic samples. The concentration dependence of pertinent ferroelectric parameters has been determined and used to develop the mathematical framework based on Ginzburg-Landau theory describing the evolution of these properties across the concentration space. We pose that combination of automated scanning probe microscope and combinatorial spread library approach will emerge as an efficient research paradigm to close the characterization gap in the high-throughput materials discovery. We make the data sets open to the community and hope that will stimulate other efforts to interpret and understand the physics of these systems.

## Introduction

Exploring the relationships between materials structure and properties across a large compositional and processing landscapes is crucial for understanding the fundamental mechanisms governing material behavior and leveraging this knowledge for the materials optimization and discovery. Combinatorial libraries,[1,2] which comprise multiple material compositions within binary, ternary, and even higher-dimensional compositional spaces, are a pivotal approach for rapidly synthesizing a large number of material compositions within a single process and sample. This approach has significantly benefited from materials fabrication techniques such as pulsed laser deposition and sputtering.[3]

However, the broad adoption of combinatorial methods as a part of materials discovery and optimization requires solution of several closely connected problems. This includes the capability to read the structural and functional properties along the composition spaces, build the physical models for materials behaviors based on these studies, and explore the effects of materials preparation.[4] The latter is particularly important given that properties of materials in the library form may differ from that of the single composition film and especially bulk material.

The foundational bottleneck lies in the need for rapid and efficient methods to characterize the properties and functionalities of combinatorial samples. This need, for the development of characterization techniques capable of offering rapid characterization of materials and novel approaches to extract properties vs. compositions/conditions, has been realized since the early days of combinatorial research. Traditionally, these studies have been based on parallel measurements via optical responses including reflectometry,[5] adsorption spectra, or photoluminescence.[6] However, structural, electrical, mechanical, and ferroelectric functionalities generally require sequential measurements and, in many cases, macroscopic device fabrication.[7]

Here, we demonstrate the automated Piezoresponse Force Microscopy (PFM) as a tool to perform high-throughput characterization of ferroelectric combinatorial systems. This approach relies on that the electromechanical detection on PFM is insensitive to the contact area.[8] We further develop the Ginzburg-Landau model describing the evolution of the properties across the composition space. Generally, automated PFM is positioned to become a key tool to close the characterization gap in combinatorial research.

## Results and Discussions

*Experiment*

We investigated the Sm-doped BiFeO$_3$ (Sm-BFO) combinatorial library, which comprises transition with an increase of Sm content from the ferroelectric state of pure rhombohedral BiFeO$_3$ to a nonferroelectric state of orthorhombic 20% Sm-doped BiFeO$_3$. These similar libraries have been explored by our groups previously, and corresponding structural and electron microscopy data is reported elsewhere.[9-11] We first employed PFM, operating in imaging mode, to characterize topography and ferroelectric domain structures of the Sm-BFO sample. Figure 1 (a-c) shows the topographic morphology from three representative locations with varying Sm content, where the morphology of the low Sm content side is smoother than that of high Sm content side. Figure 1 (d-f) and (g-i) show band excitation PFM amplitude and phase images, respectively, indicating distinct domain structure with varying Sm content.

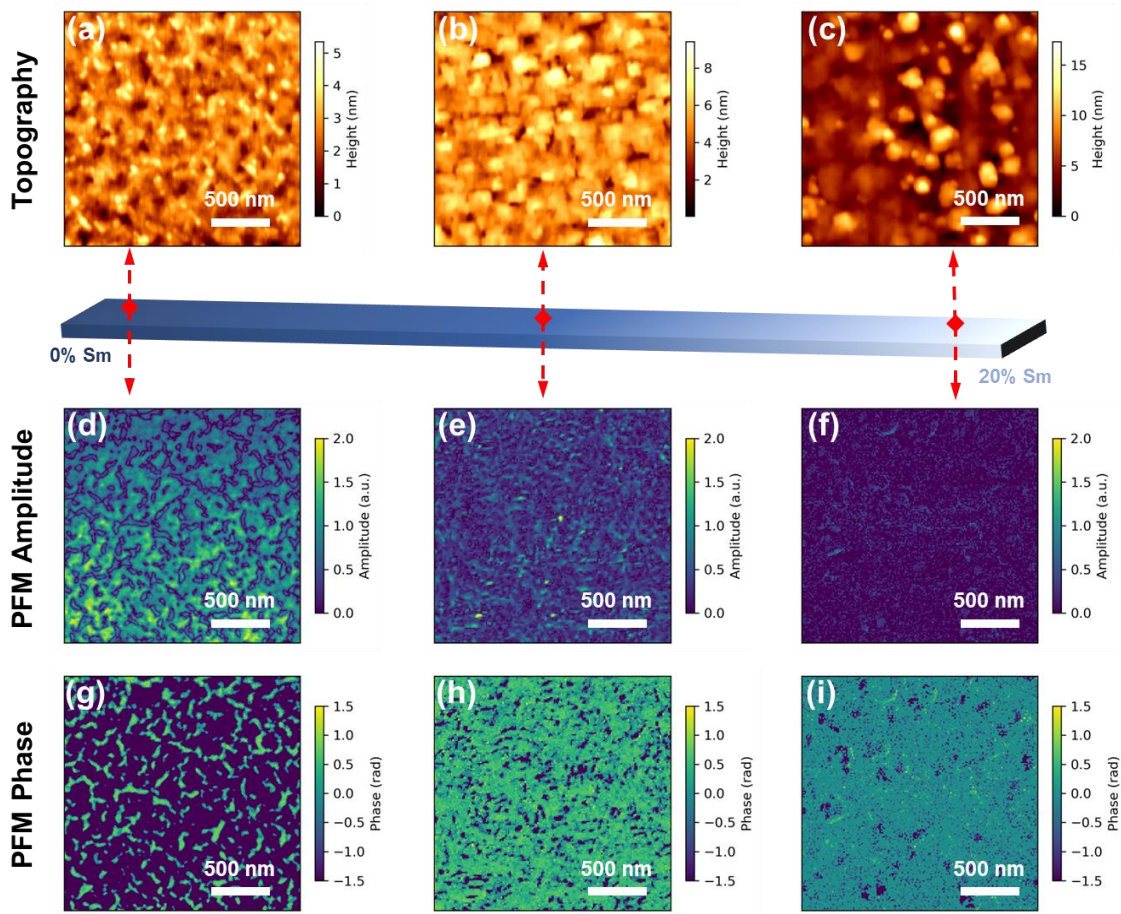

**Figure 1.** Topography and band excitation PFM image results from three representative locations corresponding to varying Sm concentration of the combinatorial Sm-BFO sample. (a-c) Topography. Band excitation PFM (d-f) amplitude and (g-i) phase images showing distinct ferroelectric domain structures from three locations with different Sm concentration.

Operating PFM in spectroscopy model further allows us to measure piezoresponse vs. voltage hysteresis loop, which is fundamental to understand ferroelectric materials. The piezoresponse vs. voltage hysteresis loop illustrates the relationship between ferroelectric polarization and applied electric field, revealing important information about remanent polarization, switching dynamics, and coercive field. PFM hysteresis loop measurements can operate at the nanoscale, enabling investigation of local variation in piezoelectric and ferroelectric properties across the sample. The important aspect is that PFM imaging and spectroscopy signal formation mechanism is very robust and is insensitive to the details of the tip-surface contact,[8,12,13] differentiating this method from other scanning probe microscopy (SPM) techniques. Similarly, the extensive theoretical apparatus have been developed to analyze PFM image formation mechanisms in terms of piezoelectric constants,[8,14,15] analyze the hysteresis loop shapes and deconvolute the domain growth mechanisms,[16,17] etc.

However, the scanner ranges of PFM are often small of the order of 10-100 µm, which fall below the dimensions of larger combinatorial libraries that are around mm-cm range. Therefore, traditional PFM measurements of combinatorial libraries must rely on human operators to manually locate different compositional regions and move the tip to the desired scan region corresponding to difference compositions.[9]

To address this challenge, we have implemented an automated PFM experiment, which combines a motorized stage in vendor microscope NanoSurf DriveAFM with AEcroscopy platform.[18,19] In our setup, the motorized stage is employed to automatically position the sample corresponding to various compositions within the combinatorial library under the PFM probe. This stage enables seamless movement along the composition spread direction without the need for manual intervention. AEcroscopy platform allows for the acquisition of hysteresis loops without human intervention.

To systematically capture hysteresis loops across the library, we have developed a Python-based workflow. This workflow performs hysteresis loop measurements at regular intervals of 100 µm along the composition spread direction; at each composition point, three hysteresis loops are acquired at positions spaced 100 nm apart along the direction perpendicular to the composition spread. To avoid scratching or other undesirable effects (e.g., sample or probe damage) that may occur during long-distance movements driven by the motorized stage, the PFM probe is withdrawn before each motorized stage movement; after reaching the next composition point, the probe is re-approached to resume measurements. Control over the precise locations for PFM hysteresis loop measurements at each composition is done by the piezo scanner, which offers better precision within a small range, ensuring that the hysteresis loops are acquired with the desired spatial resolution and alignment relative to the composition gradient. The acquisition of PFM hysteresis is done by the AEcroscopy platform, a system extensively detailed and discussed in our previous work.[18,19]

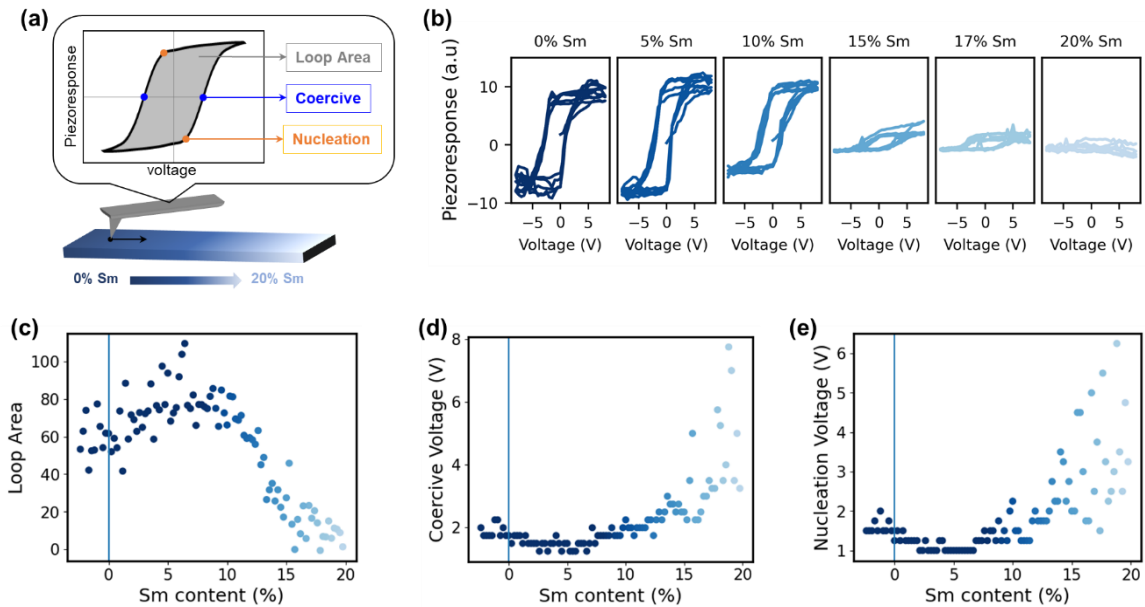

Figure 2. Automated PFM piezoresponse vs. voltage hysteresis loop measurement across the composition spread. (a), a scheme showing physical descriptors extracted from hysteresis loop. (b), a few representative hysteresis loops corresponding to different Sm content. (c-e), hysteresis loop area, coercive voltage, and nucleation voltage, respectively, as a function of Sm content. Notably, the hysteresis loops near 20% Sm are closed, so that the coercive and nucleation voltage near 20% Sm virtually lose their physical meaning. Additionally, the data points before 0% Sm also correspond to 0% Sm because there is a small range of pure BFO at the end of the combinatorial sample.

Figure 2 presents the results of PFM hysteresis loop measurements conducted across the composition spread in the combinatorial Sm-BFO system. Figure 2a depicts a standard piezoresponse vs. voltage hysteresis loop, showcasing key parameters associated with ferroelectric properties like nucleation voltage, coercive voltage, and loop area. Examining these parameters across varying Sm content can provide valuable insights into the impact of Sm concentration on the ferroelectric response. Figure 2b shows several representative hysteresis loops with different Sm contents. It is indicated that the loop diminishes as the Sm content increases, indicating a transition from a ferroelectric phase to a non-ferroelectric phase.

To gain a deeper understanding of the effect of Sm content on ferroelectricity, we extracted classical physical descriptors—loop area, coercive voltage, and nucleation voltage— from all loops.[20] The evolution of these physical descriptors as a function of Sm content is illustrated in Figure 2c-e. It is observed that the loop area increases with Sm content up to 5%, reaching a maximum around 5%-10% Sm content. Subsequently, the loop area begins to decrease beyond 10% Sm and diminishes near 20% Sm. In contrast, coercive and nucleation voltage (Figure 2d and 2e) decrease before reaching 5% Sm content and increase thereafter as a function of Sm content.

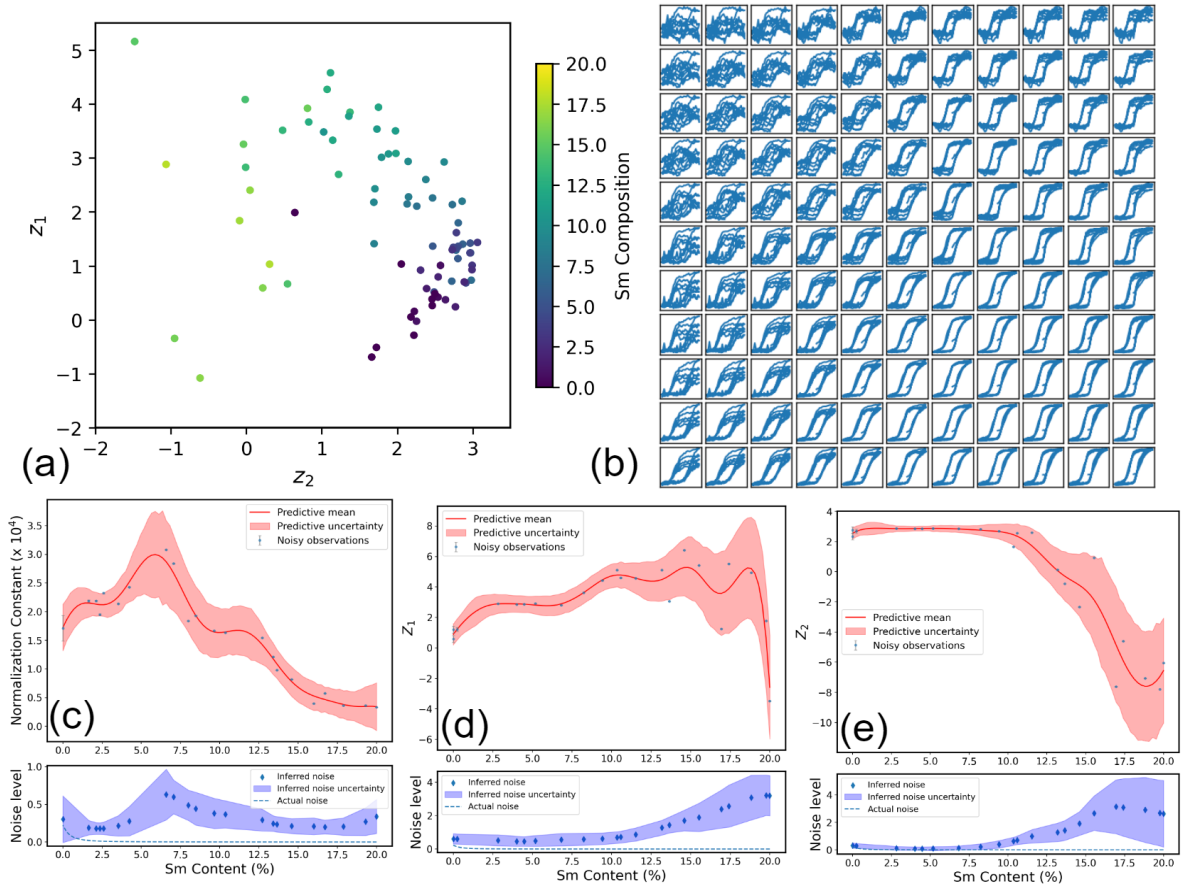

**Figure 3.** (a) Latent distribution of the normalized hysteresis loops in the latent space of the Variational Autoencoder. The colors correspond to the Sm composition. (b) corresponding latent representations of the response vs time, illustrating the characteristic traits in the data and meaning of the latent variables. (c) Composition dependence of the normalization constat, i.e. remanent polarization. (d,e) Composition dependence of the first and second latent variable. The red line corresponds to the fit by the heteroscedastic Gaussian Process, red-shaded area is associated uncertainty, and blue curves are the reconstructed noise function and its uncertainty.

The additional insights into the evolution of the hysteresis loops across the compositional space of the combinatorial library can be derived via variational autoencoder (VAE) analysis.[21] Briefly, VAE is a non-linear dimensionality method based on discovery of the optimum representation of the high-dimensional data as several low-dimensional latent variables, or latent vectors.[22] During training, the VAE aims to minimize the sum of the reconstruction loss of the data and the Kullback-Leibler divergence between the distribution of the latent variables and the Gaussian distribution $N(0,1)$. The unique property of the VAE that we use here is its capability to disentangle the representations of the data, i.e. identify and order the factors of variability inside the data set. Previously, we have extensively used the VAE for imaging data, discussing these phenomena in detail.[23-26]

Shown in Figure 3(a) is the latent distribution of the hysteresis loop data. Here, the hysteresis loops were normalized to [0,1] and time dependence of the PFM signal was used as a data set to train the VAE with the 2D latent space. With this, each spectrum is encoded via two latent variables $z_1$ and $z_2$. The color encodings of points in Figure 3(a) represents the evolution of the latent variables with the composition. Note that while the distribution is broad, the trend for shift from lower-right corner to top-right corner and then to the left of the latent space with concentration is clearly visible.

To interpret the latent variables, the latent representation of the VAE is shown in Figure 3 (b). Here, the latent representation corresponds to the data reconstructed from the rectangular grid in the latent space and provides a guide to the interpretation of latent variables. The bottom right corner is occupied by the nearly ideal hysteresis loops characteristic of a good ferroelectric. Note that the multiple loops almost coincide, suggesting nearly-perfect cyclostationary nature of switching process. On transition to the top left corner the loops maintain shape, but the difference between first and subsequent loops for positive biases becomes obvious. Shift to the left leads to the highly noisy loops.

The evolution of the latent variables across the composition space is further illustrated in Figure 3 (c-e). The normalization constant shows the increase of the response from 0 to approximately 8% Sm concentration, with rapid decrease afterwards. As expected, this behavior agrees with Figure 2 (c), and represents the enhancement of the electromechanical properties with doping when approaching the morphotropic region. It is important to note that this initial trend is almost linear, and the doping effects are felt even for small concentrations. Above the 8% region the response starts to decrease and eventually becomes zero around ~15% Sm concentration. We speculate that above this threshold the PFM hysteresis loop is dominated by the stray electrochemical contributions to the signal.[27]

Additional information can be obtained from the concentration dependence of the latent variables. Here, we note that normalization of the hysteresis data allows to explore the trends in the shape of the hysteresis loops, but also leads to the non-uniform noise distribution across the composition space. To account for this effect, we provide both the original data and the fit by the heteroscedastic Gaussian Process, estimating the trends in data and noise simultaneously. The first latent variable shows relatively weak increase across the composition space, with the drastic increase of noise after the transition.

Comparatively, the second variable shows several clear inflection points. The first corresponds to the 8% Sm concentration, where $z_2$ starts to decrease rapidly. The corresponding change in the loop shape can be interpreted from the latent representation in Figure 3 (b), corresponding to the strong change in the position of the first corresponding to subsequent loops, i.e. emergence of the non-stationary PFM behavior. We argue that this behavior is due to the emergence of the nanoscale domains that have different response compared to the structure emerging under cyclostationary conditions. This effect becomes more pronounced with the increase in Sm concentration. Finally, at 15% percent doping, the behavior changes again, presumably due to disappearance of polarization-related hysteresis and onset of the pure electrochemical behavior. Note that associated noise, as seen on the bottom panel of Figure 3(e), also shows the inflection point.

To summarize, the automated PFM experiments have revealed the systematic evolution of the polarization switching in the concentration space of the Sm-doped BFO compositional library. Both the physics-based and VAE analyses reveal the presence of MPB boundary with maximum response around 8% and transition to non-ferroelectric phase around 15%. The evolution of relevant parameters within each domain is close to linear. The VAE analysis allows to disentangle complex factors of variation within the data, complementary to the physics-based analysis.

*Theoretical model*

For the description of the perovskite solid solution $Sm_xBi_{1-x}FeO_3$ polar properties and phase diagrams we use the model of four sublattices, shortly FSM.[28] The model allows the analytical description of cation displacements $A_i$ in the four sublattices ($i$=1-4) of ferroelectric-antiferrodistortive perovskites of the $BiFeO_3$-type and explain the coexistence of the ferroelectric (FE) rhombohedral (R) phase, antiferroelectric (AFE) or/and paraelectric (PE) orthorhombic (O) phases, and ferrielectric (FEI) and/or spatially-modulated mixed (R + O) ordered phases. The FSM approach also allows to model the phase diagrams of $Sm_xBi_{1-x}FeO_3$ thin films related with the case of the Sm/Bi cation sublattice.[29]

The FSM reduces the description of the FE (R), FEI (R+O), and AFE/PE (O) phases to the thermodynamic analyses of the Landau-type free energy of the $Sm_xBi_{1-x}FeO_3$ with several dimensionless phenomenological parameters. The main parameters of the FSM are sublattices linear stiffness, sublattices coupling strength, and the gradient energy coefficients. The FSM is consistent with the Landau-Ginzburg-Devonshire (LGD) approach proposed in Ref.[30] for the pristine BiFeO$_3$ and BiFeO$_3$ slightly doped by the rare-earth elements. For the consistency, the four cation displacements should be gathered into the two order parameters: the long-range polar ordering and the structural antipolar ordering. In result we obtained the "reduced" FSM. Notably that the 2 order parameters is the minimal amount for the description of multiferroics with the coupled long-range orders.[31]

In Appendix A we consider the Landau free energy expansions with 4-th and 6-th maximal powers of the polar and antipolar order parameters corresponding to the reduced FSM. The cases are designated as 2-4 Landau free energy and 2-4-6 Landau free energy hereinafter. Appeared that the phase diagram of a bulk Sm$_x$Bi$_{1-x}$FeO$_3$ can be described by the 2-4 Landau free energy, and we need to consider the 2-4-6 Landau free energy for the adequate description of the Sm$_x$Bi$_{1-x}$FeO$_3$ thin films.

According to Refs. [28,29], one could write the effective 2-4-6 Landau free energy in the following dimensionless form:

$$G_{Effective} = \frac{1}{2}\alpha(T,x)P^2 + \frac{\beta}{4}P^4 + \frac{\gamma}{6}P^6 + \frac{1}{2}\eta(T,x)A^2 + \frac{\xi}{4}A^4 + \frac{\delta}{2}P^2A^2 - PE. \quad (1)$$

Here $P = \frac{A_1+A_2+A_3+A_4}{2}$ and $A = \frac{A_1-A_2+A_3-A_4}{2}$ are the dimensionless polar and antipolar order parameters respectively; $E$ is the dimensionless electric field component, coupled to the polarization $P$. Here $\alpha(T,x)$ and $\eta(T,x)$ are the linear stiffnesses, $\beta$, $\gamma$ and $\xi$ are the nonlinear stiffnesses, $\delta$ is the coupling strength of the polar and antipolar order parameters.

The temperature and Sm-content dependences of the dimensionless coefficients $\alpha(T,x)$ and $\eta(T,x)$ in Eq.(1) were chosen in such a way that to reproduce the experimentally observed[32] phase diagram of a bulk Sm$_x$Bi$_{1-x}$FeO$_3$:

$$\alpha(T,x) = \alpha_T\left[\frac{T}{T_C} - f_P\left(\frac{x}{x_C}\right)\right], \quad (2a)$$

$$\eta(T,x) = \eta_T \exp\left[\frac{T}{T_A} - f_A\left(\frac{x}{x_A}\right)\right]. \quad (2b)$$

The functional form Eq.(2) was selected using the following speculations. In accordance with the Landau theory, the coefficients $\alpha(T,x)$ and $\eta(T,x)$ should be linear functions of the temperature $T$, which change their sign at the characteristic temperatures. In particular, for the pure BiFeO$_3$ $\alpha(T,0)$ changes its sign at the Curie temperature $T_C$, and $\eta(T,0)$ changes its sign at the Neel temperature $T_A$. Since the Sm doping deteriorates the polar properties and improve the antipolar ordering of Sm$_x$Bi$_{1-x}$FeO$_3$, the inequalities $\alpha(T,x) > 0$ for $T > T_C$ and $\eta(T, x) > 0$ for $T > T_A$ should be valid for $0 \leq x \leq 1$. Also, the inequality $T_C \geq T_A$ should be valid at least for small $x$.

Thus, we use the trial functions (2), where $\alpha_T$ and $\eta_T$ are positive coefficients; $f_P$ and $f_A$ are positive functions of $x$, $f_P(0) = f_A(0) = 1$ and $f_{P,A}(0)$ monotonically decreases with $x$ increase. The values $x_C$ and $x_A$ are characteristic (or critical) concentrations of Sm, above which the functions $f_P$ and $f_A$ decrease strongly.

To reproduce the experimental phase diagram of a bulk $Sm_xBi_{1-x}FeO_3$,[32] we tried linear, higher polynomial-type, stretch exponential-type functions for $f_P$ and $f_A$, and lead to the conclusion that the stretch exponential-type and Gaussian functions

$$f_P\left(\frac{x}{x_C}\right) = \exp\left[-\left(\frac{x}{x_C}\right)^4\right] \text{ and } f_A\left(\frac{x}{x_A}\right) = \exp\left[-\left(\frac{x}{x_A}\right)^2\right], \qquad (3)$$

which are relatively simple, can reproduce the experimental results.[32] At that the fourth and second powers in the exponents, $\left(\frac{x}{x_C}\right)^4$ and $\left(\frac{x}{x_A}\right)^2$, reflect the differences between more slow and faster x-dependences of the polar and antipolar orderings, respectively. The values, $\alpha_T$ and $\eta_T$, $T_C$ and $T_A$, $x_C$ and $x_A$ are fitting parameters, which can vary in a reasonable range for different the 2-4 and 2-4-6 LGD models, as well as to be dependent on the sample preparation method.

Without loss of generality that one can put $\alpha_T = 1$ and $\beta = \xi = 1$ in the dimensionless LGD free energy (1) and vary the dimensionless parameter $\delta$ in the range $\left\{-\frac{1}{2}, 1\right\}$. Thus, seven fitting parameters, $\eta_T$, $T_C$ and $T_A$, $x_C$, $x_A$, $\gamma$ and $\delta$ remain.

*Phase diagram and polar properties of a bulk $Sm_xBi_{1-x}FeO_3$*

Phase diagram of a bulk $Sm_xBi_{1-x}FeO_3$ calculated using the 2-4-LGD free energy (1) and expressions (2) is shown in **Figure 4(a)**. The fitting parameters, which correspond to the best agreement with experimental results [**Error! Bookmark not defined.**], are: $T_C = 1100$ K, $T_A = 800$ K, $x_C = 0.1$, $x_A = 0.15$, $\gamma = 0$ and $\eta_T = 0.1$, and δ=0.1. The experimental phase diagram is shown in the inset to **Figure 4(a).** The FE phase corresponds to $P \neq 0$ and $A = 0$; and the AFE phase corresponds to $A \neq 0$ and $P = 0$. The mixed FEI phase corresponds to $P \neq 0$ and $A \neq 0$; and the PE phase corresponds to $A = 0$ and $P = 0$.

The coercive field and the area of the quasi-static polarization loops, which correspond to the single-domain polarization reversal, are shown in **Figure 4(b)** and **4(c)**, respectively. monotonically decreases with x increase and disappears at $x = x_{cr} \approx 0.105$. Quasistatic single-domain hysteresis loops of polarization $P$ and "anti-polarization" $A$ calculated in the external electric field is shown in **Figure 4(d)** and **4(e)**, respectively. Different loops correspond to increasing Sm-content x, which vary from 0 to 20 %. As expected, the increase of x causes the decrease of the polarization loop width, height, and its eventual disappearance. At the same conditions the two peaks of the antipolar order parameter approach one another and eventually split.

However, the 2-4-LGD model, which well fits to a bulk $Sm_xBi_{1-x}FeO_3$ do not explain the results observed experimentally by PFM in the $Sm_xBi_{1-x}FeO_3$ thin films.

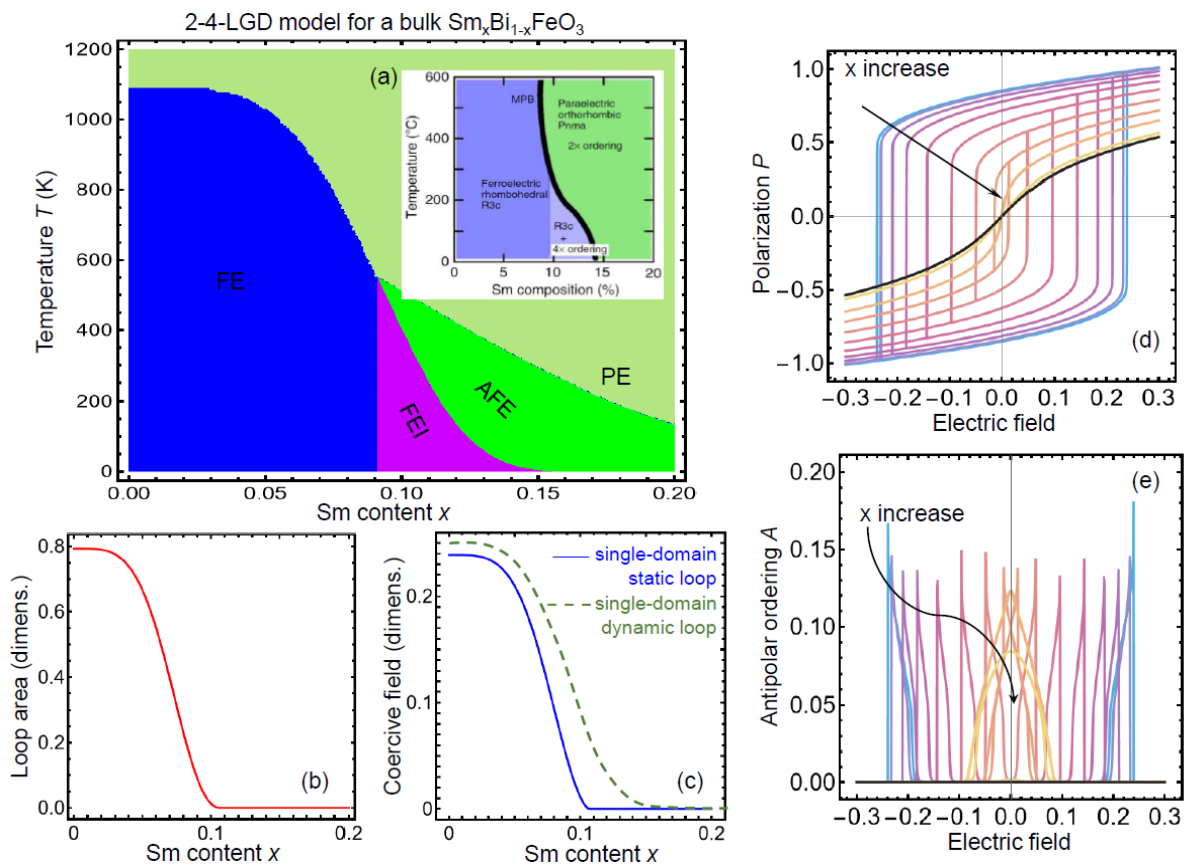

**Figure 4.** (a) Phase diagram of a bulk $Sm_xBi_{1-x}FeO_3$ calculated using 2-4-LGD free energy. Sm-content x-dependence of the area[32] (b) and coercive field (c) of the hysteresis loops. Solid curves are results for the static hysteresis loops and the dashed curve represents the results for the dynamic loops. Electric field hysteresis of polarization $P$ (d) and anti-polarization $A$ (e) calculated for $T = 300$ K and x=0, 1, 2, 3 … and 20 % (from violet to red curves). The fitting parameters are: $T_C = 1100$ K, $T_A = 800$ K, $x_C = 0.1$, $x_A = 0.15$, $\gamma = 0$ and $\eta_T = 0.1$, and δ=0.1.

*Phase diagram and polar properties of a $Sm_xBi_{1-x}FeO_3$ thin film*

Due to elastic strains and various kind of defects the phase diagram of $Sm_xBi_{(1-x)}FeO_3$ thin films can be different from the diagram of a bulk material.[29] The situation is typical for many other ferroelectric films, where the elastic strains, point and topological defects can influence strongly on the phase boundaries due to the second order and higher order electrostriction coupling.[33,34] In particular, the order of the phase transition and phase coexistence can be changed strongly.[29] The idea to select the "effective" parameters in the LGD 2-4-6 free energy, which are different from the bulk parameters, allows much better agreement with experimentally measured characteristics of the local hysteresis loops.

Phase diagram of a $Sm_xBi_{1-x}FeO_3$ thin film calculated using the 2-4-6-LGD free energy (1) and expressions (2) is shown in **Figure 5(a)**. The fitting parameters are: $T_C = 1100$ K, $T_A = 1100$ K, $x_C = 0.2$, $x_A = 0.22$, and $\eta_T = 1$, and, $\gamma = 1$, and $\delta = 3(0.05 - x)$. The fitting parameters correspond to the best agreement with experimentally measured by PFM off-field local piezoresponse hysteresis loops.

The loop area and coercive voltage, calculated in a quasistatic regime, are shown by solid curves in **Figure 5(b)** and **5(c),** respectively. The coercive field and the area of polarization loop at the first increases, pass through the diffuse maximum at $x \approx 0.1$ and then decreases with x increase. The loop area does not vanish even above 20% of Sm, because the AFE-type double hysteresis loops and/or strongly pinched hysteresis loops appear in the x-range for the quasistatic and dynamic regimes, respectively. The quasistatic coercive field, corresponding to the single hysteresis loop, disappears at $x \approx 0.22$, but the dynamic coercive field strongly increases for the same x-values. The calculated behavior of the coercive field and loop area is in a semi-quantitative agreement with the experimental results shown in **Figure 2**.

The most interesting experimental result is the diffuse maxima observed for the loop area and negative runs of nucleation and coercive voltages at about 10% of Sm content. In accordance with the calculated phase diagram, shown in **Figure 5(a)**, and corresponding parameters of hysteresis loops, shown in **Figure 5(b)** and **5(c)**, the experimental observation can be explained by appearance of the mixed FEI phase in a wide x-range, and the coexistence of FE and AFE orderings inside the FEI phase region creates the MPB-like region with the maximal loop area, remanent polarization and coercive in a definite x-range.

Corresponding hysteresis of static polarization *P* and "anti-polarization" *A* calculated in the external electric field is shown in **Figure 5(d)** and **5(e)**, respectively**.** Different loops correspond to increasing Sm-content x, which vary from 0 to 20 %. The quasistatic loop area, remanent polarization and coercive at the first increases, pass through the diffuse maximum at $x \approx 0.1$ and then decreases with x increase [see **Figure 5(d)**]. At the same conditions the two peaks of the antipolar order parameter approach one another, then slightly repulse, and eventually split [see **Figure 5(e)**].

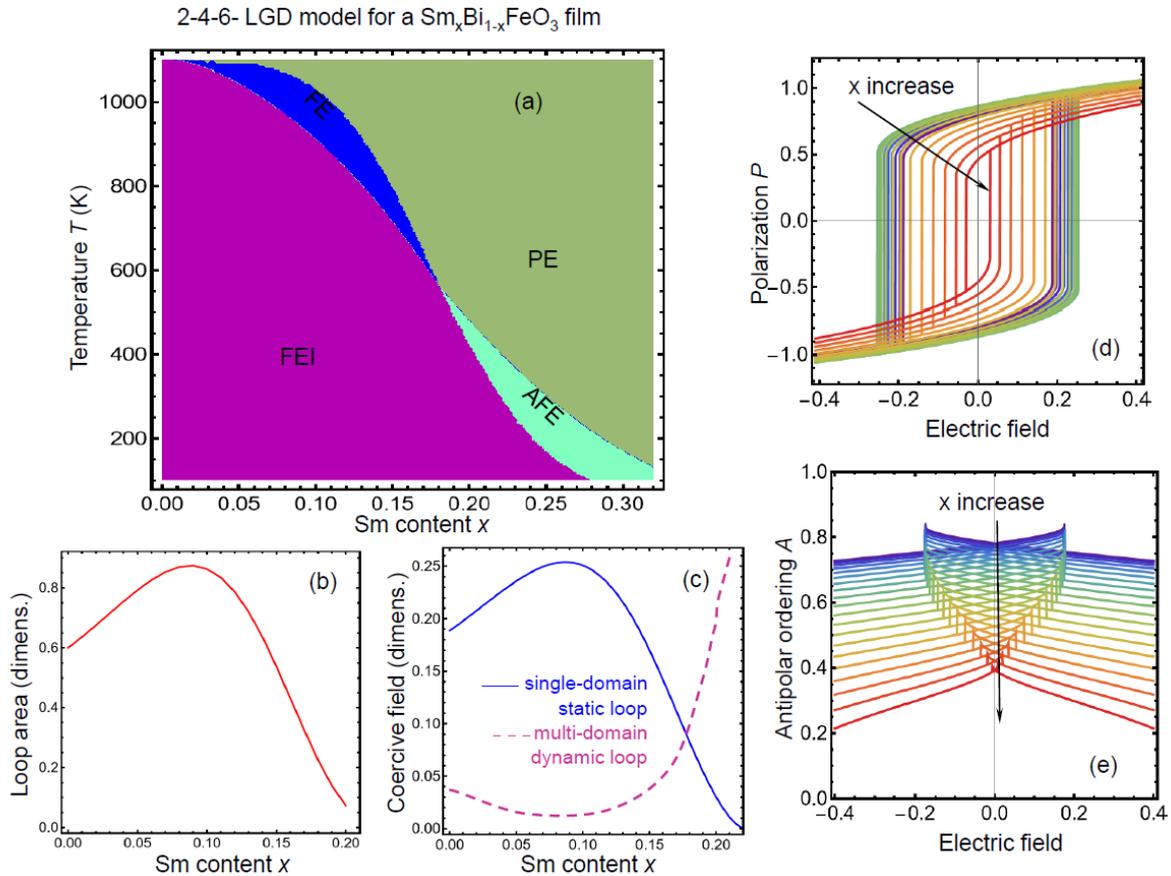

**Figure 5. (a)** Phase diagram of a $Sm_xBi_{1-x}FeO_3$ thin film calculated using 2-4-6-LGD free energy. Sm-content x-dependence of the area **(b)** and coercive field **(c)** of the hysteresis loops. Solid curves are results for the static hysteresis loops and the dashed curve represents the results for the dynamic loops. Electric field hysteresis of polarization $P$ **(d)** and anti-polarization $A$ **(e)** calculated for $T = 300$ K and x=0, 1, 2, 3 … and 20 % (from violet to red curves). The fitting parameters are: $T_C = 1100$ K, $T_A = 1100$ K, $x_C = 0.2$, $x_A = 0.22$, and $\eta_T = 1$, and $\gamma = 1$ and $\delta = 3(0.05 - x)$.

To resume, the phase diagram of a bulk $Sm_xBi_{1-x}FeO_3$ can be described by the 2-4 Landau free energy for the polar and antipolar long-range orders, but we need to consider the 2-4-6 Landau free energy for the polar and antipolar long-range orders for the adequate description of the x-dependences of the polarization hysteresis loop area observed $Sm_xBi_{1-}$
14

$_x$FeO$_3$ thin films. We also need to consider the domain formation to describe the observed x-dependence of the coercive field. At that the phase diagram of the Sm$_x$Bi$_{1-x}$FeO$_3$ thin films is different from the bulk diagram due to the several well-known reasons, such as elastic strains, point and topological defects, which result into the different fitting parameters in the 2-4-6 Landau free energy for the films. The film diagram contains a big region of the FEI phase and small (and thin) regions of the AFE and FE phases. The bulk diagram contains the big region of the FE phase and much smaller regions of the FEI and AFE phases. The increase of the FEI region is also related to the coexistence of FE and AFE orders in the morphotropic region.

**Summary**

To summarize, we have explored the evolution of the ferroelectric properties in the combinatorial Sm$_x$Bi$_{1-x}$FeO$_3$ library across the morphotropic transition between ferroelectric rhombohedral and antiferroelectric orthorhombic phases. We discovered that the polarization reaches its maximum at ~8% Sm and begins to decrease beyond 10% Sm. The hysteresis loop diminishes significantly at ~ 20% Sm, suggesting non-ferroelectricity. Conversely, the coercive and nucleation voltage decrease up to 5% Sm content, but increase as a function of Sm content thereafter.

To describe observed behaviors, we have explored a broad range of possible Landau-type models. The phase diagram of a bulk Sm$_x$Bi$_{1-x}$FeO$_3$ can be described by the 2-4 Landau free energy for the polar and antipolar long-range orders, but we need to consider the 2-4-6 Landau free energy for the polar and antipolar long-range orders for the adequate description of the x-dependences of the polarization hysteresis loop area observed Sm$_x$Bi$_{1-x}$FeO$_3$ thin films.

The phase diagram of the Sm$_x$Bi$_{1-x}$FeO$_3$ thin films is different from the bulk diagram due to the several well-known reasons, such as elastic strains, point and topological defects, which result into the different fitting parameters in the 2-4-6 Landau free energy for the films. The film diagram contains a big region of the FEI phase and small thin regions of the AFE and FE phases. The bulk diagram contains the big region of the FE phase and much smaller regions of the FEI and AFE phases. The increase of the FEI region is also related to the coexistence of FE and AFE orders in the morphotropic region.

This study suggests that the intrinsic quantitativeness of certain microscopy measurements can allow their broad use to close the characterization gap in the combinatorial research, closing the loop from synthesis to characterization. This approach will benefit from broader introduction of multifunctional SPM platforms, SPM calibration in terms of local responses, and development of quantifications routines for SPM measurements in terms of physically-relevant materials functionalities. The second necessary component will be the development of machine learning methods for accelerated exploration of combinatorial libraires since the grid approach is prohibitively slow even for the binary cases. The early attempt at this has been reported in Ref. [9]. Finally, we note that machine learning methods can further be used to systematize the discovery of underpinning physics from observational data, e.g. via the tree search/optimization across the possible spaces of free energy. Jointly, these developments will lay a new paradigm for high-throughput materials discovery and optimization.

## APPENDIX A. Theoretical description

**Table II.** Anisotropic free energy. Three components

| Phase | Spontaneous order parameters | Free energy $f_R$ |
|---|---|---|
| PE | $P_1 = P_2 = P_3 = 0$ | 0 |
| FEa | $P_1 = \pm\sqrt{-\frac{a_1}{2b_{11}}},\ P_2 = P_3 = 0$ | $-\frac{a_1^2}{4b_{11}}$ |
| FEc | $P_3 = \pm\sqrt{-\frac{a_3}{2b_{33}}},\ P_1 = P_2 = 0$ | $-\frac{a_3^2}{4b_{33}}$ |
| FEaa | $P_1 = \mp P_2 = \pm\sqrt{-\frac{a_1}{2b_{11}+b_{12}}}, P_3 = 0$ | $-\frac{a_1^2}{(2b_{11}+b_{12})}$ |
| FEac | $P_1 = \pm\sqrt{-\frac{2b_{33}a_1 - b_{13}a_3}{4b_{11}b_{33}-b_{13}^2}}, P_3 = \pm\sqrt{-\frac{2b_{11}a_3 - b_{13}a_1}{4b_{11}b_{33}-b_{13}^2}}, P_2 = 0$ | $\frac{-b_{33}a_1^2 - b_{11}a_3^2 + b_{13}a_1 a_3}{(4b_{11}b_{33}-b_{13}^2)}$ |
| FEr | $P_1 = \mp P_2 = \frac{\pm\sqrt{a_3 b_{13} - 2a_1 b_{33}}}{\sqrt{2(2b_{11}+b_{12})b_{33} - 2b_{13}^2}},$ $P_3 = \pm\frac{\sqrt{-a_3(2b_{11}+b_{12}) + 2a_1 b_{13}}}{\sqrt{2(2b_{11}+b_{12})b_{33} - 2b_{13}^2}}$ | $\frac{-a_3^2(2b_{11}+b_{12}) - 4a_1^2 b_{33} + 4a_1 a_3 b_{13}}{4\left((2b_{11}+b_{12})b_{33} - b_{13}^2\right)}$ |

Stability matrix and corresponding conditions:

$$\begin{bmatrix} \alpha + 3\beta P^2 + 5\gamma P^4 + \delta A^2 & 2\delta PA \\ 2\delta PA & \eta + 3\xi A^2 + \delta P^2 \end{bmatrix}$$

$\alpha + 3\beta P^2 + 5\gamma P^4 > 0, \eta + 3\xi A^2 + \delta P^2 > 0$

and $(\alpha + 3\beta P^2 + 5\gamma P^4)(\eta + 3\xi A^2 + \delta P^2) - 4\delta^2 P^2 A^2 > 0$

Theses ones should be considered with equations of states

$$\alpha P + \beta P^3 + \gamma P^5 + \delta P A^2 = E$$

$$\eta + \xi A^2 + \delta P^2 = 0$$

**Acknowledgements:**

The PFM characterization and analysis (Y.L.) were conducted at the Center for Nanophase Materials Sciences (CNMS), which is a US Department of Energy, Office of Science User Facility at Oak Ridge National Laboratory. This effort was supported (SVK and AR, data analysis and acquisitions) by the center for 3D Ferroelectric Microelectronics (3DFeM), an Energy Frontier Research Center funded by the U.S. Department of Energy (DOE), Office of Science, Basic Energy Sciences under Award Number DE-SC0021118. E.A.E. and A.N.M. acknowledge (theory development) the DOE Software Project on "Computational Mesoscale Science and Open Software for Quantum Materials", under Award Number DE-SC0020145 as part of the Computational Materials Sciences Program of US Department of Energy, Office of Science, Basic Energy Sciences. The work at the University of Maryland was supported by ONR MURI N00014172661, NIST cooperative agreement 70NANB17H301, and DTRA CB11400 MAGNETO, Univ. of Maryland.

**Conflict of Interest**

The authors declare no conflict of interest.

**Authors Contribution**

A.R. performed data analysis with help from Y.L. and S.V.K.. Y.L. acquired PFM data using AEcroscopy package and prepared analysis notebooks. R.P. and I.T. prepared the combinatorial sample. E.A.E. and A.N.M. performed theory analysis. M.C. assisted in NanoSurf PFM experiment. M.A.Z. developed VAE and GP. All authors contributed to discussions and the final manuscript.

**Data Availability Statement**

The data acquired in this study and associated analysis workflows are open and is freely available at https://github.com/yongtaoliu/Combinatorial-Search-in-Nanosurf.

**References**

1  Takeuchi, I., Lauterbach, J. & Fasolka, M. J. Combinatorial materials synthesis. *Materials today* **8**, 18-26 (2005).
2  Koinuma, H. & Takeuchi, I. Combinatorial solid-state chemistry of inorganic materials. *Nature materials* **3**, 429-438 (2004).
3  Green, M. L., Takeuchi, I. & Hattrick-Simpers, J. R. Applications of high throughput (combinatorial) methodologies to electronic, magnetic, optical, and energy-related materials. *Journal of Applied Physics* **113** (2013).
4  Kusne, A. G. *et al.* On-the-fly closed-loop materials discovery via Bayesian active learning. *Nature communications* **11**, 5966 (2020).
5  Schenck, P. K., Bassim, N. D., Otani, M., Oguchi, H. & Green, M. L. Design and spectroscopic reflectometry characterization of pulsed laser deposition combinatorial libraries. *Applied surface science* **254**, 781-784 (2007).
6  Wang, J. *et al.* Identification of a blue photoluminescent composite material from a combinatorial library. *Science* **279**, 1712-1714 (1998).

7       Kan, D., Long, C. J., Steinmetz, C., Lofland, S. E. & Takeuchi, I. Combinatorial search of structural transitions: Systematic investigation of morphotropic phase boundaries in chemically substituted BiFeO3. *Journal of Materials Research* **27**, 2691-2704 (2012).

8       Eliseev, E. A., Kalinin, S. V., Jesse, S., Bravina, S. L. & Morozovska, A. N. Electromechanical detection in scanning probe microscopy: Tip models and materials contrast. *Journal of Applied Physics* **102** (2007).

9       Ziatdinov, M. A. *et al.* Hypothesis learning in automated experiment: application to combinatorial materials libraries. *Advanced Materials* **34**, 2201345 (2022).

10      Nelson, C. T. *et al.* Deep learning ferroelectric polarization distributions from STEM data via with and without atom finding. *npj Computational Materials* **7**, 149 (2021).

11      Slautin, B. N. *et al.* Multimodal co-orchestration for exploring structure-property relationships in combinatorial libraries via multi-task Bayesian optimization. *arXiv preprint arXiv:2402.02198* (2024).

12      Kalinin, S. V., Karapetian, E. & Kachanov, M. Nanoelectromechanics of piezoresponse force microscopy. *Physical Review B* **70**, 184101 (2004). https://doi.org:10.1103/PhysRevB.70.184101

13      Kalinin, S. V., Shin, J., Kachanov, M., Karapetian, E. & Baddorf, A. P. in *Ferroelectric Thin Films Xii* Vol. 784 *Materials Research Society Symposium Proceedings* (eds S. HoffmannEifert *et al.*) 43-48 (2004).

14      Kalinin, S. V., Eliseev, E. A. & Morozovska, A. N. Materials contrast in piezoresponse force microscopy. *Applied Physics Letters* **88** (2006). https://doi.org:Artn 232904

10.1063/1.2206992

15      Felten, F., Schneider, G. A., Saldaña, J. M. & Kalinin, S. V. Modeling and measurement of surface displacements in BaTiO3 bulk material in piezoresponse force microscopy. *Journal of Applied Physics* **96**, 563-568 (2004). https://doi.org:10.1063/1.1758316

16      Morozovska, A. N., Eliseev, E. A. & Kalinin, S. V. Domain nucleation and hysteresis loop shape in piezoresponse force spectroscopy. *Applied Physics Letters* **89** (2006). https://doi.org:Artn 192901

10.1063/1.2378526

17      Bdikin, I. K. *et al.* Domain dynamics in piezoresponse force spectroscopy: Quantitative deconvolution and hysteresis loop fine structure. *Applied Physics Letters* **92** (2008). https://doi.org:10.1063/1.2919792

18      Liu, Y. *et al.* AEcroscoPy: A software-hardware framework empowering microscopy toward automated and autonomous experimentation. *Small Methods*, 2301740 (2024).

19      Liu, Y. *AEcroscoPy*, 2023).

20      Jesse, S., Lee, H. N. & Kalinin, S. V. Quantitative mapping of switching behavior in piezoresponse force microscopy. *Review of scientific instruments* **77** (2006).

21      Liu, Y. *et al.* Decoding the shift-invariant data: applications for band-excitation scanning probe microscopy. *Machine Learning: Science and Technology* **2**, 045028 (2021).

22      Kingma, D. P. & Welling, M. Auto-encoding variational bayes. *arXiv preprint arXiv:1312.6114* (2013).

23      Valleti, M., Liu, Y. & Kalinin, S. Physics and chemistry from parsimonious representations: image analysis via invariant variational autoencoders. *arXiv preprint arXiv:2303.18236* (2023).

24      Kalinin, S. V., Steffes, J. J., Liu, Y., Huey, B. D. & Ziatdinov, M. Disentangling ferroelectric domain wall geometries and pathways in dynamic piezoresponse force microscopy via unsupervised machine learning. *Nanotechnology* **33**, 055707 (2021).

25      Kalinin, S. V. *et al.* Deep Bayesian local crystallography. *npj Computational Materials* **7**, 181 (2021).

26	Kalinin, S. V., Dyck, O., Jesse, S. & Ziatdinov, M. Exploring order parameters and dynamic processes in disordered systems via variational autoencoders. *Science Advances* **7**, eabd5084 (2021).

27	Vasudevan, R. K., Balke, N., Maksymovych, P., Jesse, S. & Kalinin, S. V. Ferroelectric or non-ferroelectric: Why so many materials exhibit "ferroelectricity" on the nanoscale. *Applied Physics Reviews* **4** (2017).

28	Morozovska, A. N., Eliseev, E. A., Chen, D., Nelson, C. T. & Kalinin, S. V. Building a free-energy functional from atomically resolved imaging: Atomic-scale phenomena in La-doped BiFe O 3. *Physical Review B* **99**, 195440 (2019).

29	Jinling Zhou, H.-H. H., Shunsuke Kobayashi, Shintaro Yasui, Ke Wang, Eugene A. Eliseev, and Anna N. Morozovska, Zijian Hong, Daniel Sando, Qi Zhang and Nagarajan Valanoor. An Emergent Quadruple Phase in doped Bismuth Ferrite Thin Films through Site and Strain Engineering. *unpublished* (2024).

30	Karpinsky, D. V. *et al.* Thermodynamic potential and phase diagram for multiferroic bismuth ferrite (BiFeO 3). *npj Computational Materials* **3**, 20 (2017).

31	Balashova, E. & Tagantsev, A. Polarization response of crystals with structural and ferroelectric instabilities. *Physical Review B* **48**, 9979 (1993).

32	Borisevich, A. Y. *et al.* Atomic-scale evolution of modulated phases at the ferroelectric–antiferroelectric morphotropic phase boundary controlled by flexoelectric interaction. *Nature communications* **3**, 775 (2012).

33	Pertsev, N., Zembilgotov, A. & Tagantsev, A. Effect of mechanical boundary conditions on phase diagrams of epitaxial ferroelectric thin films. *Physical review letters* **80**, 1988 (1998).

34	Kvasov, A. & Tagantsev, A. K. Role of high-order electromechanical coupling terms in thermodynamics of ferroelectric thin films. *Physical Review B* **87**, 184101 (2013).